\documentclass[11pt]{article}
\usepackage{amsmath,latexsym,amssymb,graphicx}
\usepackage{appendix}
\newcommand{\beq}{\begin{equation}}
\newcommand{\eeq}{\end{equation}}

\catcode`@=11
\long\def\@caption#1[#2]#3{\par\addcontentsline{\csname
  ext@#1\endcsname}{#1}{\protect\numberline{\csname
  the#1\endcsname}{\ignorespaces #2}}\begingroup
    \small
    \@parboxrestore
    \@makecaption{\csname fnum@#1\endcsname}{\ignorespaces #3}\par
  \endgroup}
\catcode`@=12
\def\lsim{\lower3pt\hbox{$\, \buildrel < \over \sim \, $}}  
\def\gsim{\lower3pt\hbox{$\, \buildrel > \over \sim \, $}}  

\newcommand{\ba}{\begin{array}} 
\newcommand{\ea}{\end{array}}
\newcommand{\bal}{\begin{align}}
\newcommand{\eal}{\end{align}}
\newcommand{\bea}{\begin{eqnarray}}
\newcommand{\eea}{\end{eqnarray}}

\renewcommand{\lim}[2]{\begin{tabular}{c} \vspace{-3mm}lim\\ $_{#1\to #2}$\end{tabular}}

\def\ie{{\it i.e.}}

\def\gev{\,\mbox{GeV}}
\def\tev{\,\mbox{TeV}}
\def\sfrac#1#2{{\textstyle\frac{#1}{#2}}}
\def\mueff{\mu_{\rm eff}}
\def\bmueff{B_{\mu,{\rm eff}}}
\def\mpl{M_{\rm Pl}}
\def\vev#1{\left\langle #1 \right\rangle}


\begin{document}
\begin{titlepage}  

\vskip 2cm
\begin{flushright}
FERMILAB-PUB-11-615-T
\end{flushright}
\begin{center}  
\vspace{0.5cm} \Large {\sc Solving the Hierarchy Problem with a Light Singlet and Supersymmetric Mass Terms}
 %\Large {\sc A possible origin of neutrino masses.}                            
\vspace*{1.5cm}
  
\normalsize  
  
{\bf Antonio Delgado$^1$},
%~\footnote[1]{antonio.delgado@nd.edu} 
{\bf Christopher Kolda$^1$},
%~\footnote[2]{ckolda@nd.edu}
and {\bf 
Alejandro de la Puente$^{1,2}$
%~\footnote[3]{adelapue@nd.edu}
}

\smallskip   
\medskip   
{\it $^1$ Department of Physics, University of Notre Dame,}\\ 
{\it Notre Dame, IN 46556, USA}\\ ~~\\
{\it $^2$ Fermi National Accelerator Laboratory, P.O. Box 500, Batavia, IL 60510, USA}

\vskip0.6in \end{center}  
   
\centerline{\large\bf Abstract} 
\vspace{.5cm}

\noindent

A generalization of the Next-to-Minimal Supersymmetric Model (NMSSM) is studied in which an explicit $\mu$-term as well as a small supersymmetric mass term for the singlet superfield are incorporated. We study the possibility of raising the Standard Model-like Higgs mass at tree level through its mixing with a light, mostly-singlet, CP-even scalar. We are able to generate Higgs boson masses up to 145 GeV with top squarks below 1.1 TeV and without the need to fine tune parameters in the scalar potential. This model yields light singlet-like scalars and pseudoscalars passing all collider constraints. 
   
\vspace*{2mm}   
%\smallskip\newline  
  
\end{titlepage}

%\section{}
A long-standing problem for supersymmetric (SUSY) extensions of the Standard Model is how to evade the LEP bound on the mass of the lightest (Standard Model-like) Higgs boson without introducing new sources of fine-tuning into the theory. Within the Minimal SUSY Standard Model (MSSM), the lightest Higgs must lie below the $Z^0$ at tree level, and can only be pushed above the $114\gev$ reported by LEP~\cite{LEP} with top squark masses and mixings that appear to reintroduce (albeit in a small way) the hierarchy problems that SUSY is supposed to solve in the first place. Even when extending the Higgs sector of the MSSM in non-minimal directions, the mass of the lightest Higgs is tied inexorably to the $Z^0$ mass times dimensionless couplings; imposing perturbativity on those couplings up to very high scales limits their sizes and so typically preserves the little hierarchy problem. 

The prototype for non-minimal SUSY is the Next-to-Minimal SUSY Standard Model (NMSSM)~\cite{NMSSM}, which introduces just one Higgs singlet with responsibility for generating the $\mu$-terms of the MSSM dynamically. The NMSSM is best defined through its superpotential:
\begin{equation}
W=W_{Yukawa}+\lambda SH_{u}H_{d}+\sfrac{1}{3}S^{3}
\end{equation}
where $S$ is the new singlet. Once $S$ obtains a vacuum expectation value (vev), which happens naturally at the scale of electroweak symmetry breaking ($v_s\sim m_W$), a $\mu$-term naturally arises: $\mu=\lambda v_s$. In the so-called Higgs decoupling limit (in which the mass of the pseudoscalar Higgs, $A^0$, goes to infinity), the mass of the Standard Model-like Higgs boson, $h^0$, receives a positive contribution through the new $F$-term, $F_S$:
\begin{equation}
m_{h^0}\approx m_{Z^0}^{2}\cos^{2}2\beta +\lambda^{2}v^{2}\sin^{2}2\beta,
\label{nmssmresult}
\end{equation}
where $v$ is the electroweak vev ($v=174\gev$) and $\tan\beta$ is the ratio of the vevs of the usual $H_u$ and $H_d$ doublets.

As already mentioned, such a theory does not typically alleviate the little hierarchy problem much. However it is known that there are regions of parameter space within the NMSSM that {\it do}\/ solve the little hierarchy problem, at the cost of other fine tunings that must be enforced~\cite{Dermisek:2005ar}. In particular, the NMSSM only solves the little hierarchy problem if the SUSY soft-breaking terms can be balanced against the induced $\mu$-term in such a way as to sharply suppress mixing between the singlet and the lightest component of the Higgs doublets. We would argue that the underlying problem is in requiring the singlet field of the NMSSM to solve both the $\mu$-problem and the little hierarchy problem at the same time.

In a recent paper~\cite{smssm1}, we generalized the NMSSM so as to make the solution to the little hierarchy problem more natural, at least within the confines of the low-energy theory itself. Essentially we decoupled the tasks of solving the $\mu$-problem from that of raising the mass of the lightest Higgs above the LEP bound. Whereas the original NMSSM contains only dimensionless parameters within its superpotential, our version, which we called the S-MSSM (``Singlet-extended MSSM"), allows for explicit $\mu$-terms as well as explicit mass terms for the new singlet:
\begin{equation}
W=(\mu+\lambda S)H_{u}H_{d}+\frac{1}{2}\mu_{s}S^{2}.
\end{equation}
In the S-MSSM, we took the mass of the singlet, $\mu_s$, to be quite large (typically $1-3$ TeV), which suppresses $v_s$ and the mixing of the singlet with $h^0$. The singlet vev is then too small to generate the required $\mu$-term and so we included an explicit $\mu$-term as in the MSSM. Of course, in the limit that $\mu_s\to\infty$, the singlet completely decouples and we reproduce the MSSM. But our surprising result was that within the large range $\mu_s\sim 1-3\tev$ the singlet $F$-term was still large enough to raise the $h^0$ mass as high as $140\gev$ (with top squarks below a TeV), while $v_s$ was too small to mix the singlet into the $h^0$ and thereby pull the mass back down. In a follow-up paper~\cite{smssm2}, we examined the parameter space of this model within a gauge-mediated scenario to make sure that we were not introducing other sources of fine-tuning by our choices of low-energy parameters.

We want to emphasize that our previous work relied on $v_s$ being quite small, often around a GeV or so. Here two effects compete to increase or decrease the light Higgs mass. Because the $F$-term contribution of the NMSSM, Eq.~(\ref{nmssmresult}), assumes $\mu_s=0$, it is replaced by an equivalent expression in the S-MSSM that goes to zero as $\mu_s\to\infty$. But the parameter $\mu_s$ ultimately controls the size of $v_s$, specifically, $v_s\propto 1/\mu_s$. Thus larger $\mu_s$ correspond to smaller $v_s$ but also smaller corrections to the light Higgs mass. On the other hand, $v_s$ controls the mixing of the singlet into the light Higgs, and so as $v_s\to 0$, the mixing disappears, increasing the light Higgs mass. When one accounts for both effects, one finds a wide range of parameter space in which the contribution from $F_S$ is not yet decoupled while the effects of the mixing are insignificant, and here we found a solution to the little hierarchy problem.

In this paper we will consider an entirely different regime, one in which $\mu_s$ is quite small, and show that even in this regime one can find solutions to the little hierarchy problem, though for quite different reasons. Further, this is a region of parameter space in which the phenomenology at the LHC may be significantly richer than for the large $\mu_s$ limit of the S-MSSM. We will find that, in general,  this model predicts  two scalar states (mostly singlet) with masses bellow the SM like Higgs and, therefore, depending on the parameters new decays could exist for the Higgs which can make the discovery at the LHC quite challenging. 

%\section{Low Energy Model}
The most general superpotential that can be written for the MSSM with the addition of one gauge singlet, and that preserves $R$-parity, is:
\begin{equation}
W=W_{\rm Yukawa}+(\mu+\lambda S)H_{u}H_{d}+\sfrac{1}{2}\mu_{s}S^{2}+\sfrac{1}{3}\kappa S^{3}+\xi S.
\end{equation}
This superpotential allows $S$ to couple to the $H_uH_d$ bilinear, which will eventually generate the corrections to the masses of the Higgs bosons, but it also allows for an explicit $\mu$-term and an explicit mass term, $\mu_s$, for the $S$-superfield. We can also include a trilinear $S^3$ term; in the NMSSM, this term is required to avoid a $PQ$-symmetry which is broken at the electroweak scale, resulting in a massless pseudoscalar. Here the symmetry is broken (softly) by $\mu,\mu_s\neq0$, and so the $S^3$ terms plays little role except to stabilize the potential far from the origin; for the purposes of this analysis, we simply set $\kappa$ to zero.

In the presence of non-zero $\mu,\mu_s$-terms, $S$ is a true singlet and we cannot prevent the SUSY-preserving tadpole term, $W\sim\xi S$, nor a SUSY-breaking tadpole in our scalar potential, $V\sim \xi' S$. Because the $Z_3$ symmetry that is usually associated with the NMSSM superpotential is broken only softly by the explicit mass terms, we know that $\xi\sim\mu_{(s)}\Lambda$ and $\xi'\sim M_{_{\rm SUSY}}\mu_{(s)}\Lambda$, both suppressed by some power of $16\pi^2$. Here $\Lambda$ is a cutoff beyond which $S$ fails to transform as a true singlet. If $S$ is a true singlet all the way to the Planck scale, then presumably $\Lambda\simeq \mpl$, in which case $v_s$ will become quite large and destabilize the electroweak hierarchy. (One can see this explicitly if we allow $S$ to couple to the hidden sector through non-minimal K\"ahler terms, for example.) However, we will not argue here that $S$ is a singlet all the way to $\mpl$, but rather we will treat the S-MSSM superpotential as simply a low-energy effective theory valid below some cutoff $\Lambda$ which we take to be sufficiently far above $M_{_{\rm SUSY}}$ as to allow our analysis to be sensible.

For the reasons given above, we drop both the tadpole term and the cubic self-interaction, leaving:
\begin{equation}
W=W_{\rm Yukawa}+(\mu+\lambda S)H_{u}H_{d}+\frac{1}{2}\mu_{s}S^{2}.
\end{equation}
We refer to the model described by this superpotential as the S-MSSM.
Despite this being the same superpotential studied in our previous papers~\cite{smssm1,smssm2}, the analysis here will differ in an important way. In our previous analyses, it was assumed that $\mu_s$ was the largest mass scale in the (low-energy) theory, typically a few TeV. Here we will assume the opposite, namely that $\mu_s\ll \lambda v<v$. We will see that this leads to a strikingly different Higgs spectrum, yet one that can naturally evade the LEP bound on the Higgs mass and therefore solve the little hierarchy problem.

A couple comments are in order about this superpotential in the small $\mu_s$ limit. Several papers have studied a singlet-extended MSSM in the so-called PQ limit. This is the limit in which the model possesses an explicit PQ symmetry which is broken by some unknown, high-scale physics, leaving behind a mass for the would-be axion but little else. These models generate $\mu$ solely through the vev of $S$ and have no $\mu_s$ term, and in return have an extremely light axion (actually, the pseudoscalar component of the $S$-field). Refs.~\cite{PQ,schuster,barbieri} specifically studies the limit in which $\mu=\mu_s\simeq 0$. Here we are studying the same class of models, but with the PQ-breaking soft mass terms larger, though still suppressed compared to the weak scale. We will give an example of a Frogatt-Nielsen-inspired implementation of this limit later.

We begin by studying the spectrum of this model. Starting from our superpotential and adding all the allowed soft SUSY-breaking term, the Higgs potential is given by
\begin{eqnarray}
V&=&(m^{2}_{H_{u}}+|\mu+\lambda S|^{2})|H_{u}|^{2}+(m^{2}_{H_{d}}+|\mu+\lambda S|^{2})|H_{d}|^{2}+(m_{s}^{2}+\mu_{s}^{2})|S|^{2} \nonumber \\ 
&+&[B_{s}S^{2}+(\lambda\mu_{s}S^{\dagger}+B_{\mu}+\lambda A_{\lambda}S)H_{u}H_{d}+h.c.] +\lambda^{2}|H_{u}H_{d}|^{2} \nonumber \\
&+&\frac{1}{8}(g^{2}+g'^{2})(|H_{u}|^{2}-|H_{d}|^{2})^{2}+\frac{1}{2}g^{2}|H_{u}^{\dagger}H_{d}|^{2}, 
\end{eqnarray}
where $m_{s}^2$, $B_{s}$ and $A_{\lambda}$ are the soft breaking contributions associated with the singlet. Minimization of the scalar potential yields the following three conditions, analogous to those found in the MSSM:
\begin{equation}
\frac{1}{2}m_{Z}^{2}=\frac{m_{H_{d}}^{2}-m_{H_{u}}^{2}\tan^{2}\beta}{\tan^{2}\beta-1}-\mu_{\rm eff}^{2},
\end{equation}
\begin{equation}
\sin2\beta=\frac{2B_{\mu,\rm eff}}{m^2_{H_{u}}+m^2_{H_{d}}+2\mu_{\rm eff}^2+\lambda^{2}v^{2}},
\end{equation}
\begin{equation}
\lambda v_{s}=\frac{\lambda^2 v^{2}}{2}\frac{(\mu_{s}+A_{\lambda})\sin2\beta-2\mu}{\lambda^{2}v^{2}+\mu^{2}_{s}+m^{2}_{s}+2B_{s}}, \label{eq:vs}
\end{equation}
where
%where $v_{s}=\left<S\right>$ and $v^{2}=v^{2}_{u}+v^{2}_{d}$; and 
\begin{eqnarray}
\mu_{\rm eff}&=&\mu+\lambda v_{s}, \\
B_{\mu,\rm eff}&=&B_{\mu}+\lambda v_{s}(\mu_{s}+A_{\lambda}).
\end{eqnarray}
We will be considering the case in which $\mu_s$ is small compared to the weak scale: $\mu_s\ll v$. For now let us also consider the limit in which $m_s^2$ and $B_s$ are also small compared to $v^2$; the explicit formulae simplify significantly, and we will put $m_s^2$ and $B_s$ back in for our numerical studies. In this limit, Eq.~(\ref{eq:vs}) simplifies greatly:
\beq
\lambda v_s\simeq \sfrac12 A_\lambda \sin2\beta - \mu
\eeq
which immediately leads to the surprising result that
\beq
\mueff\simeq \sfrac12 A_\lambda \sin2\beta,
\eeq
which is independent of $\mu$! That is, for small $\mu_s$, $B_s$ and $m_s^2$, the vev of $S$ aligns in such a way as to cancel the explicit $\mu$-term completely, leaving an effective $\mu$-term which is due entirely to $A_\lambda$. Meanwhile the effective $B_\mu$ term becomes:
\bea
\bmueff&\simeq& B_\mu+\sfrac12 A_\lambda^2\sin2\beta -\mu A_\lambda\\
&\simeq& B_\mu + A_\lambda (\mueff-\mu) \nonumber
\label{eq:bmueff}
\eea
which, unlike $\mueff$, does depend on the explicit $\mu$-term.

In the absence of explicit CP-violating phases in the Higgs sector, the physical spectrum of this model includes a single charged Higgs boson ($H^{\pm}$), three neutral scalars which we label \{$h_s,h,H$\}, and two neutral pseudoscalars \{$A_{s},A$\}. The states labelled with the subscript will turn out to be dominantly singlet states, while the non-subscripted states have only a small singlet component and therefore resemble their eponymous MSSM cousins. 

For the state most resembling the usual pseudoscalar Higgs, the mass is generated as in the MSSM:
$$m_A^2 = \frac{2\bmueff}{\sin 2\beta}+\cdots$$
where $\bmueff$ is given in Eq.~(\ref{eq:bmueff}) above. The ellipsis represents terms which are small compared to the weak scale, except when $A^2_\lambda\gg B_\mu,\mu^2$, in which case the leading correction is simply $\delta m_A^2 = \lambda^2 v^2$. Note that we can arrange, by proper choice of $B_\mu$, $A_\lambda$ and $\mu$ to have $\mueff\sim O(m_Z)$ while $m_A\ll m_Z$. In this way we can arrange for our model to reproduce the parameter space studied by Dermisek and Gunion~\cite{Dermisek:2005ar} in which the Higgs boson lies below the LEP bound but escapes detection by decaying dominantly into $h^0\to A^0A^0$, with $A^0$ below the threshold for decay into a pair of $b$-quarks. However, though this limit does exist, we do not see it as particularly natural or likely in this model.

In order to identify the mass eigenstates of the scalar Higgs bosons, we must diagonalize a symmetric $3\times 3$ mass matrix. It is helpful in this case to forgo the usual $\{H_d,H_u,S\}$ basis and instead rotate the upper $2\times2$ submatrix by the angle $\beta$, thereby working in the basis of 
$\{H_d\cos\beta+H_u\sin\beta,H_u\cos\beta-H_d\sin\beta,S\}$. This is the basis in which the upper $2\times2$ submatrix of the $3\times3$ {\it pseudoscalar}\/ mass matrix is diagonalized, or equivalently, the basis in which the scalar masses of the MSSM are diagonalized in the large $m_A$ limit. In this basis, the mass matrix has a simple form:
\beq
M_H^2 = \left(\begin{array}{ccc} m_Z^2\cos^2 2\beta + \lambda^2 v^2\sin^2 2\beta & (m_Z^2-\lambda^2 v^2)\sin 2\beta \cos 2\beta & 0 \\
& m_A^2 + (m_Z^2-\lambda^2 v^2)\sin^2 2\beta & \lambda v A_\lambda \cos 2\beta \\
& & \lambda^2 v^2  \end{array}\right) \label{massmatrix}
\eeq
One observes that, in this basis and in the limit of $\mu_s,m_s^2,B_s\ll \lambda^2 v^2$, there is no mixing of the singlet into the lighter MSSM-like Higgs at lowest order, a fact noticed already in Ref.~\cite{barbieri}. In fact, in the large $m_A$ limit, the mixing vanishes entirely. Yet, the light Higgs (\ie, the (1,1) element) still receives the same contribution to its mass from $F_S$ that it picks up in the NMSSM. To leading order in $m_Z^2/m_A^2$, the light Higgs mass is simply:
\beq
m_h^2 \simeq m_Z^2\cos^2 2\beta + \lambda^2 v^2 \sin^2 2\beta - \frac{(m_Z^2 - \lambda^2 v^2)^2}{m_A^2}\sin^2 2\beta \cos^2 2\beta. \label{hmass}
\eeq
The last term above represents the correction from the non-decoupling of the $A^0$, and has almost the same form as in the MSSM except that it is generically smaller than in the MSSM thanks to an expected partial cancellation between $m_Z^2$ and $\lambda^2 v^2$.

The mass of the remaining neutral, MSSM-like Higgs particle is readily derived:
\beq
m_H^2 \simeq m_A^2 + (m_Z^2-\lambda^2 v^2) \sin^2 2\beta 
+\frac{(m_Z^2-\lambda^2 v^2)^2}{m_A^2}\sin^2 2\beta\cos^2 2\beta 
- \frac{\lambda^2 v^2 A_\lambda^2}{m_A^2}\sin^2 2\beta
\eeq
%m_H^2 &=& m_A^2 + (m_Z^2-\lambda^2 v^2) \sin^2 2\beta + \frac{\lambda^2 v^2}{m_A^2}\left[4A_\lambda \mu_s- (A_\lambda+\mu_s)^2\sin^2 2\beta\right] \nonumber \\
%& & +\frac{(m_Z^2-\lambda^2 v^2)^2}{m_A^2}\sin^2 2\beta\cos^2 2\beta +\cdots\eea
where we drop terms of $O(\mu_s/\lambda v)$ and $O(m_s^2/\lambda^2 v^2)$. For $\lambda\gsim 0.5$, this state will fall just below the $A^0$ in mass.

Among the states that are mostly singlet-like, there is a scalar and a pseudoscalar:
\bea
m^2_{A_s} &\simeq& \mu_s^2 +\lambda^2 v^2 -\frac{\lambda^2 v^2 A_\lambda^2}{m_A^2}, \label{mAs}\\
m^2_{h_s} &\simeq& \mu_s^2 + \lambda^2 v^2 -\frac{\lambda^2 v^2 A_\lambda^2}{m_A^2}\cos^2 2\beta.
\eea
%m^2_{A_s} &=& \mu_s^2 +\lambda^2 v^2 -\frac{\lambda^2 v^2}{m_A^2}(A_\lambda - \mu_s)^2 +\cdots \\
%m^2_{h_s} &=& \mu_s^2 + \lambda^2 v^2 -\frac{\lambda^2 v^2}{m_A^2}\left\lbrace(A_\lambda+\mu_s)^2 \cos^2 2\beta\right. \\ \nonumber
%&& \left.+ 2A_\lambda\frac{\lambda^2 v^2 \mu_s + A_\lambda(\mu_s^2+m_s^2)}{\lambda^2 v^2} \sin^2 2\beta +4A_\lambda\mu \frac{\mu_s^2+m_s^2}{\lambda^2 v^2} \sin 2\beta\right\rbrace+\cdots
Because these states can be quite light, we have shown explicitly the effect of $\mu_s$ on their masses. Notice also that the mostly-singlet scalar is usually heavier (though only slightly) than the mostly-singlet pseudoscalar.

It is worthwhile to compare and contrast this result with the usual NMSSM in which $\mu=\mu_s=0$. In particular, it would appear that this model is hardly different, because we could take $\mu,\mu_s\to0$ and still have a sizable $\mueff=\lambda v_s$. Further, we find dynamically that $\mueff\simeq \frac12 A_\lambda \sin 2\beta$, which is exactly the relation one requires in the NMSSM to avoid large mixing of the singlet into the SM-like Higgs, namely~\cite{ellhug}
\beq
A_\lambda\simeq \frac{2\mueff}{\sin 2\beta} - 2\kappa v_s,
\eeq
in the $\kappa\ll 1$ limit.
In the usual NMSSM, this relation must hold in order to keep the mass of the SM-like Higgs boson above the LEP bound, but it must be added as an additional constraint (or tuning) on the parameters of the model; here we seem to generate it almost for free. This automatic cancellation of the singlet-doublet mixing is also present in the PQ-limit models. In both cases, 
this is due to the fact that one can allow $\kappa\to 0$ (or as small as we want) because there is no PQ axion that becomes massless as $\kappa\to 0$ in the S-MSSM. In the PQ models, this is solved by including an explicit mass for the PQ axion, but here the would-be axion gets a mass directly. (As an aside, if one sets $B_\mu=0$ in our model, then Eq.~(\ref{mAs}) simplifies to $m_{A_s} = \mu_s$ and so $\mu_s$ can be thought of as playing the role of the small PQ-breaking that occurs in Refs.~\cite{PQ,schuster,barbieri}.)

An interesting question is whether there is any ultraviolet construction that might lead naturally to the limit of our model we are studying here. Consider the PQ symmetry as it manifests itself in this model, namely with charges $+1$ for $H_u$ and $H_d$, and charge $-2$ for $S$. Under this symmetry, $\lambda$ is neutral and therefore naturally of $O(1)$. But, treated as spurions, the $\mu$-term carries charge $-2$, $\mu_s$ carries charge $4$ and $\kappa$ carries charge $6$ (and likewise for the corresponding $A$- and $B$-terms). We can then treat the low-energy model {\it \`a la}\/ Froggatt and Nielsen~\cite{Froggatt:1993di} and assume that the PQ symmetry is being broken by the vev of some field $\Theta$ with one unit of PQ-charge, and the breaking is being communicated by some heavy field(s) of mass $M$. In that case, it is natural that $\mu\gg\mu_s$ and $\kappa\ll 1$. For example, if $\theta$ carries a PQ charge of $-2$ then holomorphy and the PQ symmetry would together require:
$$\mu\sim\frac{\vev{\Theta}^2}{M}, \quad\quad\mu_s\lsim\frac{\vev{\Theta}^4}{M^4}\,m_{\rm SUSY}, \quad\quad
\kappa\lsim \frac{\vev{\Theta}^6}{M^7}\,m_{\rm SUSY}$$
where one or more powers of the scale $m_{\rm SUSY}\sim m_W$ are necessary to break the holomorphy of the superpotential. For example, one could obtain $\mu_s$ through the operator:
$$\mu_s S^2 \sim \int d^2\bar\theta\, \left(\frac{\Theta^\dagger}{M}\right)^4 \left(\frac{X^\dagger}{{\cal M}}\right)  S^2 + h.c. $$
where $X$ is the SUSY-breaking spurion, and ${\cal M}$ is the SUSY-breaking messenger scale, such that $F_X/{\cal M}\sim m_{\rm SUSY}$.
(The corresponding $A$- and $B$-terms would scale as above, multiplied by an additional power of $m_{\rm SUSY}$.) 

In this paper, we will restrict ourselves to analyzing the Higgs spectrum of this model, examining in particular whether it is possible to have a spectrum which naturally passes all current constraints. We want to especially ensure that it is possible to keep the SM-like Higgs above the LEP bound without requiring unnaturally large top squark masses or mixing. while minimizing the one loop contribution arising from the top quark and squark. We also include the dominant (and negative) 2-loop contributions to the mass of SM-like Higgs boson using FeynHiggs~\cite{Feyn}. Of particular importance is the coupling $\lambda$, which we maximize under the condition that it remain perturbative up to the apparent grand unification scale of $2\times10^{16}$ GeV; this is equivalent to setting $\lambda$ equal to it infrared pseudo-fixed point value. This leads to an upper bound on the parameter $\lambda$ which varies with $\tan\beta$, but maximizes at $\lambda\simeq 0.7$.

The parameter space of this model is quite different from the usual NMSSM. In particular, the singlet gets a vev through $A_\lambda$ and not by having $m_s^2<0$. By choosing $m_s^2>0$ in the S-MSSM we avoid any potential cosmological problems associated with run-away directions in the potential when the universe's temperature $T\sim 100\gev$~\cite{schuster}. In fact, we take $m_s^2=0$ in our analyses that follow, but will comment on non-zero values near the end. 

One important constraint on the parameter space is the LEP bound on the chargino mass, $m_{\chi^+}>94\gev$, which translates into a bound on $\mueff$: $|\mueff|>94\gev$. Assuming small $m_s^2$, this translates into a lower bound on $A_\lambda$: $A_\lambda > 190\gev/\sin2\beta$. Thus for small $\tan\beta$, $A_\lambda$ is bounded from below by roughly $190\gev$; for large $\tan\beta$ the bound on $A_\lambda$ becomes much larger, implying that the electroweak symmetry-breaking in the model is becoming fine-tuned. As we will see, even if we accepted that fine-tuning, the mass of the $h$ falls below the LEP bound for $\tan\beta\gsim 5$ (because the $S$-induced corrections go as $\sin^2 2\beta$), and so the large $\tan\beta$ region is doubly bad. Thus our model predicts that $\tan\beta$ will be small, somewhere less than 5.

We examine this model by scanning over a wide parameter space with $0\le B_{\mu}\le (1000\gev)^2$, $0\le A_{\lambda}\le 700\gev$, and $0\le\mu\le500\gev$. We keep $\mu_s$ light: $0\le\mu_{s}\le50\gev$. We also simplify the parameter space by setting $m_{s}=B_s=0$. 

\begin{figure}[ht]
\begin{center}
\includegraphics[width=0.8\linewidth]{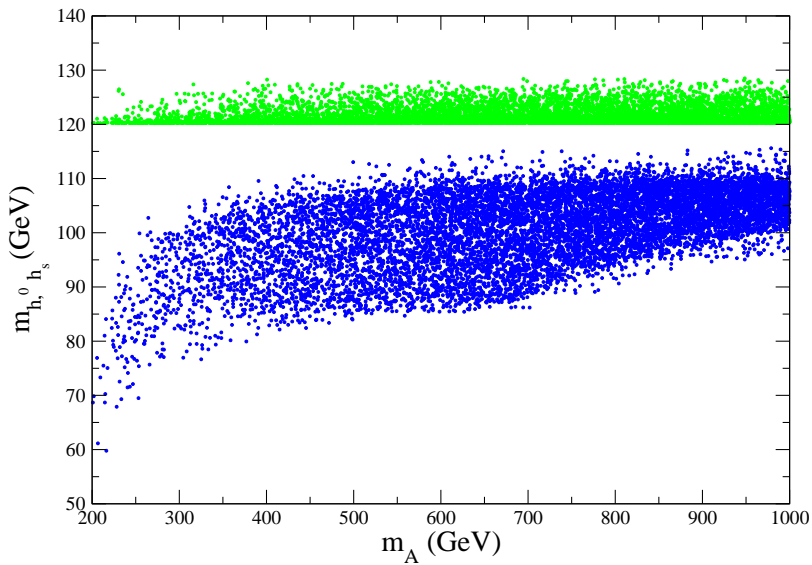}
\caption{Scatter plot of $m_{h}$ (green) and $m_{h_{s}}$ (blue) as function of $m_{A}$ with a stop mass $m_{\tilde{t}}=500\gev$ and no stop mixing. See text for additional parameters.}\label{fig:hscatter}
\end{center}
\end{figure}

\begin{figure}[ht]
\begin{center}
\includegraphics[width=0.8\linewidth]{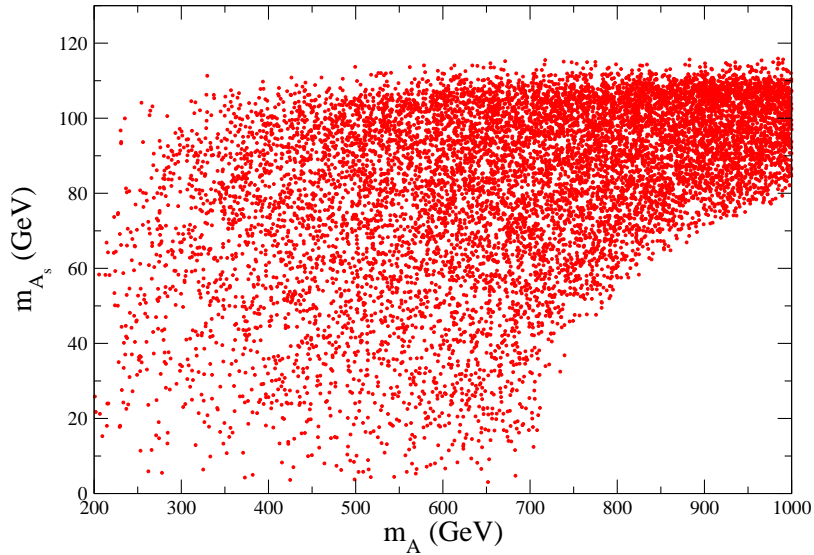}
\caption{Scatter plot of $m_{A_{s}}$ as function of $m_{A}$ using same parameter set as in Fig.~\ref{fig:hscatter}. }\label{fig:ascatter}
\end{center}
\end{figure}

In Figure~\ref{fig:hscatter} we show the masses of the SM-like Higgs, $h$, and the mostly-singlet scalar, $h_s$, as a function of the MSSM-like pseudoscalar mass, $m_A$ for a sample of models with $\tan\beta=2$ and $\lambda=0.63$. For this figure we have restricted $m_{\tilde t} = 500\gev$ and taken $A_t=0$ to minimize the stop mixing. These conditions essentially represent a minimum 1-loop contribution of the top squarks to the light Higgs mass, avoiding any hint of tuning coming from the stop sector. In Figure~\ref{fig:ascatter} we show the corresponding masses of the singlet-like pseudoscalar, $A_s$. 

%%%%%%%%%%%%%%%%%%%%%%%%%%%%%%%%%%%%%%%%%%%%%%%%%%%%%%%%%%
%%%
%%%%%%%%%%%%%%%%%%%%%%%%%%%%%%%%%%%%%%%%%%%%%%%%%%%%%%%%%%

For every point in the figures, the $h_s$ and $A_s$ masses are consistent with the LEP bound, due to their small coupling to the $Z$. In principle, the region for which $m_h<114\gev$ and both $m_{h_{s},A_{s}}<2m_{b}$ is phenomenologically viable~\cite{Dermisek:2005ar}, but we don't find points that fall into that region without fine tuning the parameters in the model, and so we don't display those.

Looking at Fig.~\ref{fig:hscatter}, one sees that, apart from a few point at low $m_A$, the mass of the SM-like Higgs is bounded from below by about $118\gev$. This is easy to understand as it follows directly from Eq.~(\ref{hmass}). If $m_A$ is large, the last, negative term decouples, and the Higgs mass is bounded from below at tree level by $(m_Z^2 \cos^2 2\beta + \lambda^2 v^2 \sin^2 2\beta)^{1/2}$. To this are added one- and two-loop corrections that are nearly universal (given a constant $m_{\tilde t}$ and $A_t$). The little bit of scatter above the lower bound is due to the small corrections from the finite $m_A$, the small effects of including non-zero $\mu_s$, and non-universal one-loop corrections, including those that arise from $A_\lambda$.

One also sees from the figures that the singlet-like scalars are expected to be below the SM-like Higgs, with masses that can be as low as a few GeV. We have checked the coupling of these states to the $Z$ and their production cross-section at LEP and excluded any points at which the singlet-like scalars would have been detected. For example, we find that, among the points in the figure, the cross-section for $e^+e^-\to Zh_s$ is at least 10 times smaller than the SM cross-section for $e^+e^-\to Zh$ with $m_h\equiv m_{h_s}$, and often it is many orders smaller. 

\begin{figure}[ht]
\begin{center}
\includegraphics[width=0.8\linewidth]{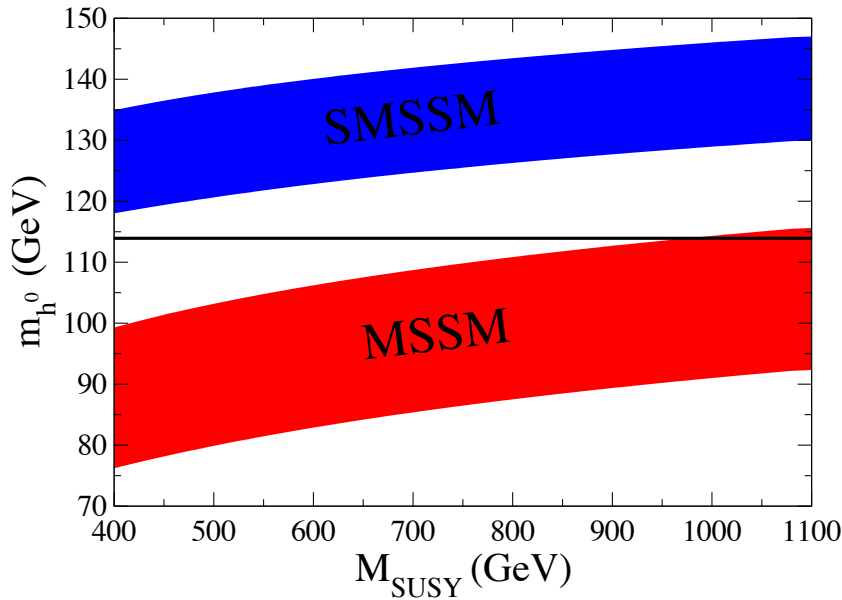}
\caption{Range of mass for $h_0$ for the S-MSSM and the MSSM  as a function of $M_{\rm SUSY}$. See the text for additional parameters.}\label{fig:compare}
\end{center}
\end{figure}

In Fig.~\ref{fig:compare}, we want to explore the effects of varying the top squark masses and their mixing. We have taken one particular choice of model in the S-MSSM: $\mu_{s}=20\gev$, $\mu=B_{\mu}=0$, $A_{\lambda}=280\gev$, and $m_{s}^2=0$ for $\tan\beta=2$ and $\lambda=0.63$. We then set the gluino and stop masses to all be equal to $M_{\rm SUSY}$ and vary them from $400$ to $1100\gev$. We also vary $A_t$ from 0 to the maximal mixing case ($A_t=\sqrt{6}\,m_{\tilde t}$), and represent this range as the upper band in the figure. In the lower band we show the MSSM for the same choices of $M_{\rm SUSY}$ and $A_t$, with $m_A\to\infty$. As can be seen from the figure for the entire range of $M_{\rm SUSY}$ the S-MSSM prediction for $m_{h^0}$ is above the LEP bound whereas for the same parameters the MSSM can only accommodate masses above the LEP bound for high masses of $M_{\rm SUSY}$. In the S-MSSM one can even have masses for the Higgs very close to the lower bound from the LHC ($\sim$ 150 GeV)~\cite{LHC}.
%%%%%%%%%%%%%%%%%%%%%%%%%%%%%%%%%%%%%%%%%%%%%%%%%%%%%%%%%%
%%%
\begin{figure}[ht]
\begin{center}
\includegraphics[width=0.8\linewidth]{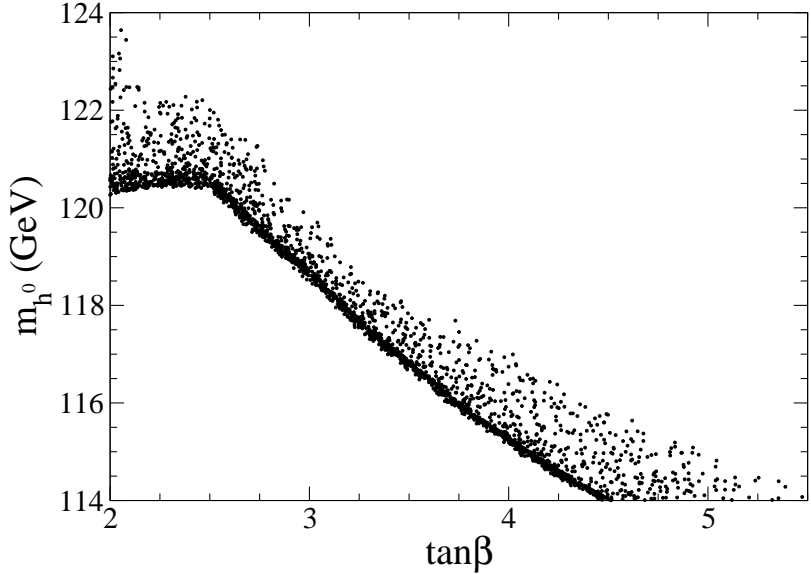}
\caption{Scattered plot for $m_h^0$ as a function of $\tan\beta$ for the same parameters as  Fig.~\ref{fig:hscatter}..}\label{fig:varytb}
\end{center}
\end{figure}
%%%%%%%%%%%%%%%%%%%%%%%%%%%%%%%%%%%%%%%%%%%%%%%%%%%%%%%%%%%

Finally, we show a plot in which we vary $\tan\beta$. In Fig.~\ref{fig:varytb} we show a scatter of random models in the same range of parameters as for Figs.~\ref{fig:hscatter}-\ref{fig:ascatter}, but now with varying $\tan\beta$ (shown along the $x$-axis), and again with $m_{\tilde t}=500\gev$ and $A_t=0$. The solid line represents a lower bound on the Higgs mass in all such S-MSSM models.
One sees immediately that the S-MSSM automatically produces SM-like Higgs bosons with masses exceeding the LEP bound for $\tan\beta\lsim 3.8$ (assuming $m_{\tilde t}> 500\gev$), with some models having sufficiently heavy Higgs masses for $\tan\beta\lsim5$. At $\tan\beta\gsim5$, the effects of the singlet on the light Higgs mass disappear (they scale as $\sin^2 2\beta\sim 1/\tan^2\beta$), and so we return to an MSSM-like spectrum at moderate to large values of $\tan\beta$. Were we to allow $m_{\tilde t}$ to increase, or to invoke larger stop mixing, the range of allowed of $\tan\beta$ would only increase.

%%%%%%%%%%%%%%%%%%%%%%%%%%%%%%%%%%%%%%%%%%%%%%%%%%%%%%%%%%
%\begin{figure}[tb]
%\begin{center}
%\includegraphics[width=0.8\linewidth]{tempo.pdf}
%\caption{Caption.}\label{fig:varytb}
%\end{center}
%\end{figure}
%%%%%%%%%%%%%%%%%%%%%%%%%%%%%%%%%%%%%%%%%%%%%%%%%%%%%%%%%%%
 
One of the simplifications we have used in examining the parameter space of the S-MSSM has been in setting $m_s^2=0$ throughout. However there is no need for this condition, and it is in fact somewhat unnatural, because there are contributions to the one-loop renormalization group equation for $m_s^2$ that are proportional to $A_\lambda^2$, and so a large $A_\lambda$ will tend to lead to an equally large $m_s^2$ unless the mediation scale for SUSY breaking is not particularly high. One finds several complications as one turns on $m_s^2$, but keeping $\epsilon=m_s^2/(\lambda^2 v^2)\ll 1$. First, the vev of $S$ shifts slightly, which causes $\mueff$ to pick up a slight dependence on the explicit $\mu$:
\bea
\lambda v_s &\simeq& \sfrac12 (1-\epsilon)\left(A_\lambda \sin2\beta - \mu\right) \\
\mueff &\simeq& \sfrac12 (1-\epsilon) A_\lambda \sin2\beta + \epsilon \mu.
\eea
More importantly, 
the exact zero in the $(1,3)$-element of the scalar Higgs mass matrix (see Eq.~(\ref{massmatrix})) is no longer zero, picking up terms that scale as $\epsilon$, and thereby inducing a mixing of the $S$-scalar into the SM-like Higgs state. However, these mixings are suppressed by powers of $m_A^2$, and so in the Higgs decoupling limit, the $S$ again decouples from the $h$:
$$\delta m_h^2 \simeq \left(\frac{m_s^2}{m_A^2}\right) 2A_\lambda \sin2\beta(A_\lambda\sin 2\beta-2\mu).$$
Notice that this contribution can have either sign, either raising or lowering the mass of $h$. Were $h$ the lightest eigenstate of the scalar mass matrix, then the mixing could only lower its mass; but here $h$ is the middle eigenstate, and so mixing with a lighter state, $s$, can actually serve to push up the mass of $h$. Either way, the effect is small as $m_A$ becomes large.

%{\it Collider Phenomenology.}\  
In our previous papers, in which we set $\mu_s\sim O(\mbox{TeV})$ in the S-MSSM, we found enhanced Higgs masses but little else for distinctive phenomenology. That was because there were no new light states, and the only real clue to the existence of the singlet was the enhanced Higgs mass itself. However in the light singlet version of the S-MSSM, we find a number of new, light states and some of these can have a profound effect on phenomenology at the LHC. 

One source for new phenomenology is the extended scalar sector, in particular the two light states $h_s$ and $A_s$. We have calculated the $ZZh_s$ coupling for all points in our parameter space, and we find it to be generically quite small, as mentioned earlier, due to the suppressed mixing between the Higgs doublets and the singlet. This is likewise true for the $ZA_sh_s$ coupling as well as couplings of the $h_s$ and $A_s$ to SM fermions, making it difficult to directly produce either at the LHC. On the other hand, there are regions of parameter space in which $h$ could decay dominantly into $h_sh_s$ or $A_sA_s$ which will mean that at the LHC the predominant decay of the Higgs will be into multijets, making its discovery quite challenging.

To summarize we have presented a singlet extension of the MSSM, the S-MSSM, in which the singlet field plays no role in the explanation of the $\mu$-problem but, on the other hand, provides a solution to the little hierarchy problem. By including supersymmetric masses for both the Higgs doublets ($\mu$) and the singlet ($\mu_s$), and then taking the limit of small $\mu_s$, we have shown that the model predicts a mass for the SM-like Higgs which are above the LEP bound for a large region of the parameter space of the model without requiring a heavy sparticle spectrum. It can even accommodate masses very close to the current LHC bound so that some regions of the parameter space of this model are going to be probed very soon by the LHC. Finally, in this model one can also find regions of parameter space in which
the main decay of the Higgs is into four jets and therefore the discovery strategies are quite different from those for the SM Higgs.

\section*{Acknowledgements} This work was partly supported by 
by the National Science Foundation under grants PHY-0905383-ARRA and PHY-0969445. ADP was supported in part by the Fermilab Fellowship in Theoretical Physics. Fermilab is operated by Fermi Research Alliance, LLC, under Contract DE-AC02-07-CH11359 with the US Department of Energy.
%%%%%%%%%%%%%%%%%%%%%%%%%%%%%%%%%%%%%%%%%%%%%%%%%%%%%%%%%%%%%%%%%%%%%%%%%

\end{document}